\documentclass[5p,sort&compress,times]{elsarticle}

\journal{Optics Communications}

\usepackage{lineno,hyperref}
\usepackage{graphicx}
\usepackage{physics}

\newcommand{\me}{\text{e}}
\newcommand{\mi}{\text{i}}

\begin{document}

\begin{frontmatter}

\title{Fock-State Superradiance in a Cold Atomic Ensemble}

\author[UFMG]{Davi~F.~Barros\corref{cor1}}	\ead{dfbarros@fisica.ufmg.br}
\author[UFPE]{Luis~F.~Mu\~{n}oz-Mart\'{\i}nez\fnref{fn1}}
\author[UFPE]{Luis~Ortiz-Guti\'{e}rrez}
\author[UdeA]{Camilo~A.~E.~Guerra}
\author[UFPE]{Johan~E.~O.~Morales}
\author[UFPE]{Raoni~S.~N.~Moreira}
\author[UFPE]{Nat\'alia~D.~Alves}
\author[UFPE]{Ayanne~F.~G.~Tieco}
\author[UFPE]{Daniel~Felinto}
\author[UFMG]{Pablo~L.~Saldanha}

\cortext[cor1]{Corresponding author}
\fntext[fn1]{Permanent address: Departamento de Ciencias B\'{a}sicas, Universidad del Sin\'{u}-El\'{\i}as Bechara Zain\'{u}m, Cra 1w No. 38--153, Monter\'{ \i}a, C\'{o}rdoba, Colombia.
}

\address[UFMG]{Departamento de F\'{\i}sica, Universidade Federal de Minas Gerais, 30161--970, Belo Horizonte --- MG, Brazil}
\address[UFPE]{Departamento de F\'{\i}sica, Universidade Federal de Pernambuco, 50670--901, Recife --- PE, Brazil}
\address[UdeA]{Instituto de F\'{\i}sica, Universidad de Antioquia UdeA, Calle 70 No. 52--21, Medell\'{\i}n, Colombia}

\begin{abstract}
A simplified theory for the wavepackets of the photons emitted during the read process of a quantum memory formed by cold atoms is provided. We arrive at analytical expressions for the single- and double-photon emissions, evidencing superradiant features in both cases. In the two-photon case, both photons are emitted in the same spatiotemporal mode, characterizing a superradiant emission of a Fock state of light with two excitations. Experiments confirm the theoretical predictions with a satisfactory agreement.
\end{abstract}


\end{frontmatter}

\section{Introduction}

Superradiance is a collective phenomenon that affects the emission of photons from an ensemble of atoms, causing it to be faster than what would be observed if each atom emits independently \cite{PhysRev.93.99, GROSS1982301}. Despite the existence of classical models for superradiance \cite{GROSS1982301}, at the single photon level, it is a genuine quantum effect, with no classical analog. Single-photon superradiance can be seen as a manifestation of the wave-particle duality since an interferometric process increases the emission rate of a light quantum particle.

Our group has a line of works experimentally demonstrating superradiance at the single-photon level \cite{1367-2630-15-7-075030, PhysRevA.90.023848, PhysRevLett.120.083603}. Our setup uses an atomic memory based on the Duan--Lukin--Cirac--Zoller (DLCZ) protocol~\cite{duan2001long}, observing superradiant photon emissions in the reading process of the memory. The fundamental indistinguishability about which atom from an ensemble emits the photon leads to an acceleration of the emission rate due to the interference of the different probability amplitudes. It was verified that the superradiant decay rate in the reading process of the atomic memory was proportional to the number of atoms in the ensemble \cite{PhysRevA.90.023848}, as expected. Finally, in Ref.~\cite{PhysRevLett.120.083603}, two-photon superradiance was experimentally studied for the first time to our knowledge. These results demonstrate the power of the atomic memories based on the DLCZ protocol as convenient tools for the experimental study of superradiance.

Here we develop a theory for the temporal and spatial wavepackets of the photons emitted by the quantum memory in the read stage, both for single-photon and for two-photon emissions, evidencing the system superradiance. This theory is different from the theory previously developed in Ref.~\cite{1367-2630-15-7-075030} for single-photon emissions, such that it could be extended for two-photon emissions and is more intuitive. Some of the results of this theory for two-photon emissions were used in Ref.~\cite{PhysRevLett.120.083603} to justify the experimental results. Here we compare it to many other experimental results with a satisfactory agreement. The theory also predicts that, when there is a two-photon emission, both are emitted in the same spatiotemporal mode, such that the quantum memory can be used to store a Fock state of light with multiple photons, with potential applications in quantum metrology and quantum information. We present a set of experimental results that confirm this behavior.

In Sec.~\ref{scheme} we briefly describe the DLCZ quantum memory scheme with one and two excitations stored on the memory. In Sec.~\ref{theory1} we provide the theory for the wavepacket of the photon emitted on the reading process of the memory when one excitation is stored on the memory, evidencing superradiant features. In Sec.~\ref{theory2} we extend this theory to deduce the wavepackets of the two photons that are emitted on the reading process of the memory when two excitations are stored on the memory. In Sec.~\ref{sec:experiments} we show our experimental results, that confirm the theoretical predictions with a satisfactory agreement. Finally, in Sec.~\ref{conclusions} we provide our concluding remarks.

\section{Quantum Memory Scheme}
	\label{scheme}

\begin{figure}[ht]
	\centering
	\includegraphics[scale=.25]{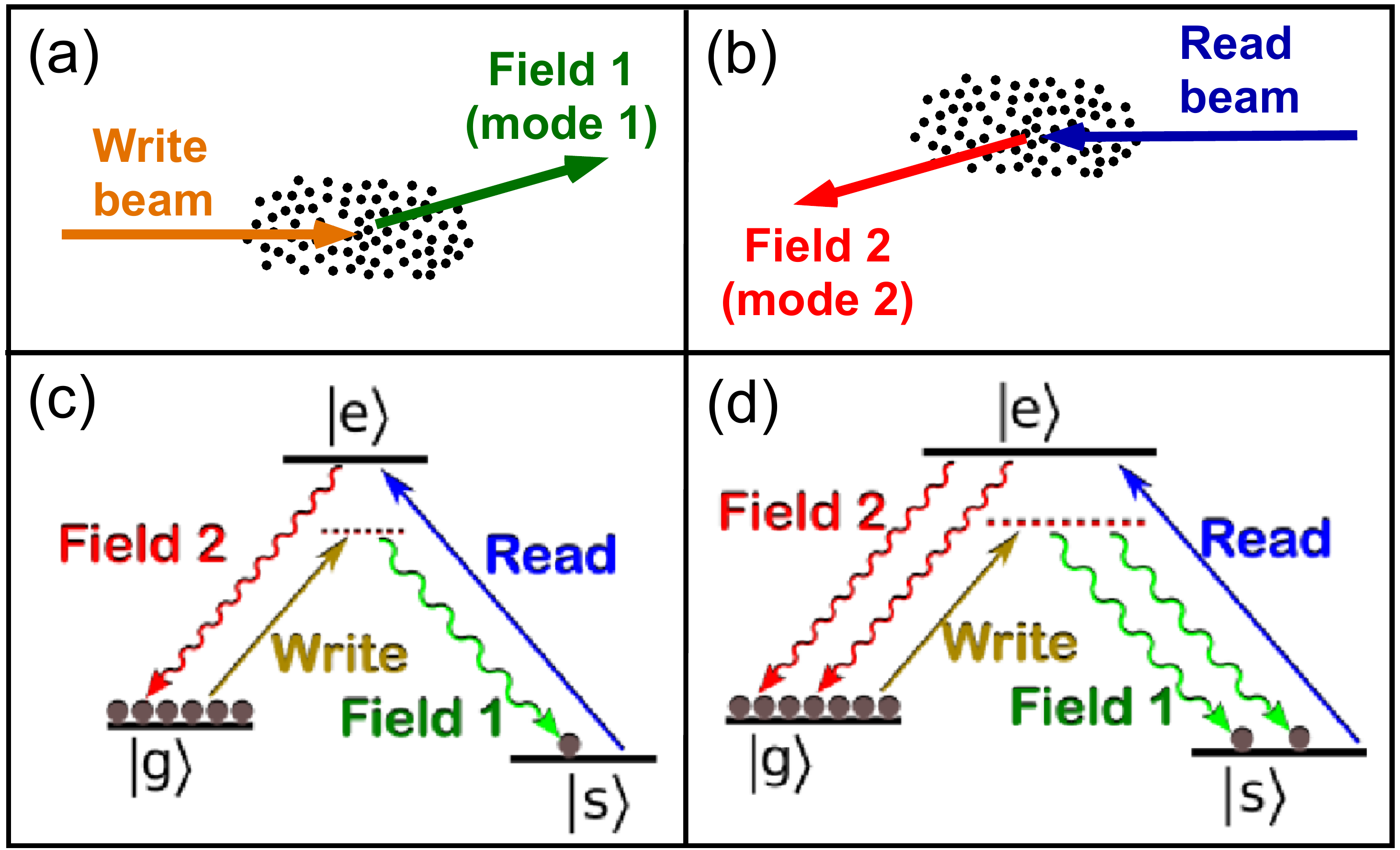}
	\caption{
	\label{sketch}
Quantum memory scheme. (a) A write laser beam may induce  $\ket{g} \rightarrow \ket{e} \rightarrow \ket{s}$ transitions on the atoms of the memory with the emission of photons into mode 1. (b) The incidence of a read laser beam converts the excitations stored on the memory into photons emitted into mode 2. (c) Representation of the writing and reading processes when one excitation is stored in the memory. (d) Representation of the writing and reading processes when two excitations are stored in the memory.}
\end{figure}

Figure \ref{sketch} illustrates the idea of the experiments we want to model, based on the DLCZ quantum memory \cite{duan2001long}. An ensemble of three-level atoms is initially prepared with all atoms on level $\ket{g}$. In its turn, a plane-wave write laser beam with wavevector $\vb{k}_\text{w}$ may perturbatively induce a $\ket{g} \rightarrow \ket{e} \rightarrow \ket{s}$ transition in one or more atoms, with the corresponding emission of one or more photons, that may be emitted into mode $1$, as depicted in Fig.~\ref{sketch}(a) . In the experiments,  mode $1$ is selected by a single-mode optical fiber, and detectors can detect up to two photons emitted into this mode during this write process on the atomic memory.
	
When only one photon is detected on the write process, due to the fundamental indistinguishability of which atom emitted the photon into mode 1, there will be a coherent superposition of which atom was left in state $\ket{s}$, with a collective atomic state that depends on the mode of the detected photon. This quantum information is stored on the atomic memory. Then, a plane-wave read laser beam with wavevector $\vb{k}_\text{r}$ resonant with the $\ket{s} \rightarrow \ket{e}$ transition allows the atom in the level $\ket{s}$ to go back to level $\ket{g}$ in a $\ket{s} \rightarrow \ket{e} \rightarrow \ket{g}$ transition with the corresponding emission of one photon into mode $2$, as depicted in Fig. \ref{sketch}(b). This mode, due to phase-matching conditions and the fundamental indeterminacy of which atom made the transitions, is the mode conjugate to mode 1, as will be shown. The whole process of writing and reading the memory in this case is depicted in Fig. \ref{sketch}(c).
	
If two photons are detected on the write process, there will be a coherent superposition of which two atoms were left in state $\ket{s}$. The read laser beam allows the atoms in level $\ket{s}$ to go back to level $\ket{g}$  with the corresponding emission of two photons into mode $2$. The whole process of writing and reading the memory in this case is depicted in Fig.~\ref{sketch}(d).

\section{Theory for Single-Photon Superradiance}
	\label{theory1}

In this section, we develop a theory for the single-photon superradiance in the quantum memory which is based on the theory developed in Ref.~\cite{1367-2630-15-7-075030}. The main difference between our treatments and previous theories that describe the behavior of quantum memories \cite{PhysRevLett.84.5094, RevModPhys.75.457, PhysRevA.73.021803} is that we assume that the read beam induces an electromagnetically induced transparency in the medium for the emitted photon \cite{RevModPhys.77.633}, removing this complicated effect from our treatments. With some further approximations, we can describe the read process of the memory as a combination of Rabi oscillations with a superradiant spontaneous emission by the atoms, arriving at an analytical expression for the wavepacket of the emitted photon. Compared to the theory of Ref.~\cite{1367-2630-15-7-075030}, the theory to be presented here is more elementary and intuitive, but less general since it considers a resonant read beam and disregards the decoherence generated by noisy magnetic fields. Nevertheless, the simplifications are consistent with our experimental conditions, and the more straightforward theory can be extended to the study of many excitations in the atomic ensemble, as we do in the following section. We do not consider the polarization of the photons in the treatment since they are not relevant to the goals of our theoretical study.

\subsection{Writing in the memory}
	\label{write}

The initial atomic ensemble state is defined as $\ket{G}$, representing a state with all atoms in the ground level $\ket{g}$. To fix the notation, let's define the operator $\sigma_{ba,i}$ as the operator that takes the state of atom $i$ from $\ket{a}$ to $\ket{b}$: $\sigma_{ba,i}\equiv \dyad{b}{a}_i$. The application of the operator $\sigma_{eg,i}$ on $\ket{G}$ leads the system to a quantum state defined as $\ket{e_i}$, where only atom $i$ is in level $\ket{e}$ and all others are in level $\ket{g}$. Similarly, the action of $\sigma_{se,i}$ on $\ket{e_i}$ drives the system to the state $\ket{s_i}$, where only atom $i$ is in level $\ket{s}$ and all others are in level $\ket{g}$.

The write beam is considered to be a plane wave with wavevector $\vb{k}_{\text{w}}$, that is detuned from the $\ket{g} \rightarrow \ket{e}$ transition. With small probability, a $\ket{g} \rightarrow \ket{e} \rightarrow \ket{s}$ transition is induced in one of the atoms of the ensemble with the emission of one photon into mode 1, as depicted in Fig. \ref{sketch}. The annihilation operator related to this mode is defined as $a_1 = \int \dd[3]{\vb{k}} \phi_1(\vb{k}) a(\vb{k})$, where $a(\vb{k})$ is the annihilation operator for a plane-wave mode with wavevector $\vb{k}$. The system's initial state is considered to be $\ket*{G} \ket*{\text{vac}}$, where $\ket*{\text{vac}}$ represents the vacuum state for the electromagnetic field of all modes orthogonal to the write beam mode. If a $\ket{g} \rightarrow \ket{e} \rightarrow \ket{s}$ transition is performed, the operator that models this is $\sum_i \int \dd[3]{\vb{k}} h(\vb{k}) \me^{\mi (\vb{k}_\text{w} - \vb{k}) \cdot \vb{r}_i} \sigma_{sg, i} a^\dag(\vb{k})$, where $h(\vb{k})$ is an interaction term that takes in consideration the intensity of the write laser \cite{1367-2630-13-7-073015}. This operator is associated to the annihilation of one photon from the write beam and the emission of one photon by one of the atoms from the ensemble, with a coherent superposition of which atom performed the $\ket{g} \rightarrow \ket{e} \rightarrow \ket{s}$ transition. The detection of the emitted photon in mode 1 will then project the electromagnetic system state on the state $a_1^\dag(\vb{r})\ket*{\mathrm{vac}}$, such that the atomic state conditioned to this detection is
\begin{equation}
	\label{1a}
	\ket{1_{a}} = \sum_i \alpha_i \ket{s_i},
\end{equation}
with
\begin{equation}
	\label{alphai}
	\alpha_i \propto \int \dd[3]{\vb{k}}' \phi_1(\vb{k}') \me^{\mi(\vb{k}_\text{w} - \vb{k}')\cdot\vb{r}_i} ,
\end{equation}
while satisfying the normalization condition $\sum_i \abs{\alpha_i}^2 = 1$. It was considered that the function $h(\vb{k})$ present in the interaction is approximately constant along the region where $\phi_1(\vb{k})$ considerable. This is a reasonable assumption, since the function $h(\vb{k})$ is associated to the pattern of an oscillating dipole emission, while $\phi_1(\vb{k})$ selects a paraxial Gaussian mode in the experiments we want to model. Eqs.~\eqref{1a} and \eqref{alphai} show that the detected photon mode is transferred to the collective atomic state. This quantum information is stored on the atomic memory until the read begins.

\subsection{Reading the memory}

In the read process, the dynamics of the atom and radiated field is described by a Hamiltonian $H(t) = H_0 + H_1(t)$. $H_0$ is the free Hamiltonian of the atoms and electromagnetic field, given by \cite{mandel1995optical}
\begin{equation}
	H_0 = \sum_i \pqty{-\hbar \omega_{se} \dyad{s}_i - \hbar \omega_{ge} \dyad{g}_i } + \int \dd[3]{\vb{k}} \hbar \omega \,a^\dag(\vb{k})a(\vb{k}),
\end{equation}
where the first term represents the atoms' energies, with $\hbar\omega_{ge} = E_e - E_g$ and $\hbar\omega_{se} = E_e - E_s$ ($E_e$ is defined as zero), and the second term represents the field's energy, with $\omega = c \abs{\vb{k}}$. $H_1(t)$ is an interaction Hamiltonian, given by
\begin{multline}
	H_1(t) = \sum_i \mi \hbar \frac{\Omega_0}{2} \me^{-\mi (\vb{k}_\text{r} \cdot \vb{r}_i - \omega_\text{r} t)} \sigma_{se,i} \\
		+ \sum_i \int \dd[3]{\vb{k}} \mi \hbar g^*(\vb{k}) \me^{-\mi \vb{k} \cdot \vb{r}_i} a^\dag(\vb{k}) \sigma_{ge,i}  + \text{H.~c.},
\end{multline}
where the first term describes the interaction of the atoms with the read beam field with angular frequency $\omega_\text{r}$ and wavevector $\vb{k}_\text{r}$, whose field amplitude $E_0$ and dipole moment of the coupling $p_{se}$ determine the frequency $\Omega_0 = p_{se} E_0/\hbar$ of the induced Rabi oscillations, and the second term describes the interaction of the atoms with the quantized electromagnetic field, that induces an spontaneous decay along the $\ket{e} \rightarrow \ket{g}$ transition on each atom \cite{mandel1995optical}. In the present analysis, we discard the fluorescence from the transition $\ket{e} \rightarrow \ket{s}$. In Ref.~\cite{1367-2630-15-7-075030} it was shown that a spontaneous photon emission from this transition destroys the coherence of the atomic state. Since we are interested in obtaining the wavepackets of the photons emitted in mode 2, this fluorescent decay causes only photon losses, with no effect in the form of these wavepackets.

In the interaction picture, the local Hamiltonian $H_0$ and $H_1(t)$ give the interaction Hamiltonian $H_{\text{int}}(t) = U_0^\dag(t) H_1(t) U_0(t)$, computed as
\begin{multline}
	\label{hint}
	H_{\text{int}}(t) = \sum_i \mi \hbar \frac{\Omega_0}{2} \me^{-\mi \vb{k}_\text{r}\cdot \vb{r}_i} \sigma_{se,i} \\
		+ \sum_i \int \dd[3]{\vb{k}} \mi \hbar g^*(\vb{k}) \me^{-\mi[\vb{k} \cdot \vb{r}_i - (\omega  - \omega_{ge}) t]} a^\dag(\vb{k}) \sigma_{ge,i} + \text{H.~c.}
\end{multline}
	Due to the resonance condition $\omega_{se} = \omega_\text{r}$, the interaction with the read beam now contributes with a time-independent operator. If the initial system state is the atomic state of Eq.~\eqref{1a} with a vacuum state for the quantized electromagnetic field, we can see that at posterior times the Hamiltonian of Eq.~\ref{hint} will generate the state
\begin{equation}
	\label{gen1}
	\ket{\psi_1(t)} = \sum_i \bqty{\alpha_i(t) \ket{s_i} + \beta_i(t) \ket{e_i}} + \int \dd[3]{\vb{k}} \gamma'(t;\vb{k}) \ket{G,\vb{k}}
\end{equation}
since its first term can induce transitions between $\ket{s_i}$	and $\ket{e_i}$ (it is implicit in the notation that these states have a vacuum state for the quantized electromagnetic field) and its second term can induce transitions between $\ket{e_i}$ and $\ket{G,\vb{k}}$ (a state with all atoms in level $\ket{g}$ and one photon with wavevector $\vb{k}$). At time $t=0$, when the read beam laser is turned on, we have $\alpha_i(0)=\alpha_i$ from Eq.~\eqref{alphai}, $\beta_i(0)=0$, $\gamma'(0;\vb{k})=0$.

By using the Schr\"odinger equation with the general state of Eq.~\eqref{gen1} subjected to the Hamiltonian of Eq.~\eqref{hint}, we obtain the following set of differential equations for the coefficients $\alpha_i(t)$, $\beta_i(t)$ and $\gamma'(t;\vb{k})$:
\begin{subequations}
	\label{schequations}
	\begin{align}
		\label{schequationsa}
		\dot{\alpha}_i(t) & = \frac{\Omega_0}{2} \me^{-\mi \vb{k}_\text{r}\cdot\vb{r}_i} \beta_i(t), \\
		\label{schequationsb} 
		\dot{\beta}_i(t) & = -\frac{\Omega_0}{2} \me^{\mi \vb{k}_\text{r}\cdot\vb{r}_i} \alpha_i(t) - \int \dd[3]{\vb{k}} g(\vb{k}) \me^{\mi[\vb{k}\cdot\vb{r}_i - (\omega - \omega_{ge})t]} \gamma'(t;\vb{k}), \\
		\label{schequationsc}
	\dot{\gamma}'(t;\vb{k}) & = \sum_i g^*(\vb{k}) \me^{-\mi[\vb{k}\cdot\vb{r}_i - (\omega  - \omega_{ge})t]} \beta_i(t),
	\end{align}
\end{subequations}
that govern the system's dynamics. To simplify the treatment, we assume that $\alpha_i(t) = \alpha(t) \alpha_i$ and $\beta_i(t) = \beta(t) \beta_i$, with $\alpha_i$ given by Eq.~\eqref{alphai}. This means that there is only one relevant spatial mode describing the ensemble's dynamics and that the functions $\alpha(t)$ and $\beta(t)$ would rule the time evolution of these collective states. The natural choice for $\beta_i$ includes the term $\me^{\mi \vb{k}_\text{r}\cdot\vb{r}_i}$ present in Eq.~\eqref{schequationsa}:
\begin{equation}
	\label{betai}
	\beta_i = \alpha_i \me^{\mi \vb{k}_\text{r}\cdot\vb{r}_i}.
\end{equation}
These collective modes can be defined as
\begin{equation}
	\label{sechi}
	\ket*{s_\chi} \equiv \sum_i \alpha_i  \ket{s_i}, \qand \ket*{e_\chi} \equiv \sum_i  \beta_i \ket{e_i}.
\end{equation}

We make similar considerations for the coefficient $\gamma'(t;\vb{k}) \equiv \gamma(t;\vb{k})\Phi(\vb{k})$. According to Eq.~\eqref{schequationsc}, the choice for $\Phi(\vb{k})$ that simplifies Eq.~\eqref{schequationsc} when written in terms of $\gamma(t;\vb{k})$ is
\begin{equation}
	\label{phi}
	\begin{split}
	\Phi(\vb{k}) & \equiv \sum_i \beta_i \me^{-\mi \vb{k}\cdot\vb{r}_i} \\
	  & \propto \sum_i \int \dd[3]{\vb{k}'} \phi_1(\vb{k}') \me^{\mi (\vb{k}_\text{w} - \vb{k}' + \vb{k}_\text{r} - \vb{k})\cdot\vb{r}_i},
	\end{split}
\end{equation}
where Eqs.~\eqref{alphai} and \eqref{betai} were used. From this equation for the coefficient $\Phi(\vb{k})$ we can deduce the directionality of the photon emitted in the read process of the memory. Since the read beam propagates in the opposite direction of the write beam, we have $\vb{k}_\text{w} + \vb{k}_\text{r} \approx 0$. Thereafter, the sum of the phases $\me^{-\mi (\vb{k} + \vb{k}')\cdot\vb{r}_i}$ in Eq.~\eqref{phi} over the positions of the very large number of atoms of the ensemble, disposed in a volume with dimensions much larger than the fields wavelenghts, is small unless $\vb{k} \approx -\vb{k}'$. So, comparing Eqs.~\eqref{alphai} and \eqref{phi}, we must have $\Phi(\vb{k})\propto \phi_1(-\vb{k})$ and the photon emitted in the read process is in the conjugate mode of the photon detected in the write process of the memory. The whole process works in a four-wave mixture configuration with the fields from the write and read processes of the memory.

With the above considerations, the quantum state of Eq.~\eqref{gen1} can be written as
\begin{equation}
	\label{psi1}
	\ket{\psi_1(t)} = \alpha(t) \ket*{s_\chi} + \beta(t) \ket*{e_\chi} + \int \dd[3]{\vb{k}} \gamma(t;\vb{k}) \Phi(\vb{k}) \ket{G,\vb{k}},
\end{equation}
and the set of differential equations in Eqs.~\eqref{schequations} {are used to get the time-evolution of $\alpha(t)$, $\beta(t)$, and $\gamma(t;\vb{k})$. To get the new differential equations we sum both sides of Eq.~\eqref{schequationsa} over all atoms noticing that, with the use of Eq.~\eqref{betai}, a term $\sum_i \alpha_i$ factorizes on both sides and can be eliminated. In Eq.~\eqref{schequationsb}, we multiply both sides of the equation by $\beta^*_i$ before summing over all atoms. Then, we can see that a term $\sum_i \abs{\beta_i}^2 = 1$ simplifies the expression on the left hand side, and a term $\sum_i \abs{\alpha_i}^2 = 1$ simplifies in the first term of the right hand side. With the use of Eq.~\eqref{phi}, the position-dependent part of the second term on the right side of Eq.~\eqref{schequationsb} can be written as $\sum_i \beta_i^* \me^{\mi\vb{k}\cdot\vb{r}_i} = \Phi^*(\vb{k})$, such that the dependency on the position disappears. Finally, we note that in Eq.~\eqref{schequationsc} a term $\sum_i \beta_i\me^{-\mi\vb{k}\cdot\vb{r}_i}$ factorizes on both sides. So the set of Eqs.~\eqref{schequations} becomes}
\begin{subequations}
	\begin{align}
		\label{newequationsa}
		\dot{\alpha}(t) & = \frac{\Omega_0}{2} \beta(t), \\
		\label{newequationsb}
		\dot{\beta}(t) & = -\frac{\Omega_0}{2} \alpha(t) - \int \dd[3]{\vb{k}} g(\vb{k}) \abs{\Phi(\vb{k})}^2 \me^{-\mi (\omega  - \omega_{ge})t} \gamma(t;\vb{k}) , \\
		\label{newequationsc}
		\dot{\gamma}(t;\vb{k}) & = g^*(\vb{k})\me^{\mi (\omega  - \omega_{ge})t} \beta(t).
	\end{align}
\end{subequations}

The principal quantity that we want to deduce from our model to compare with the experiments is the probability density $\rho_1(t)$ of the photon emission by the atomic cloud during the read process, from which we compute the probability of photodetection between times $t$ and $t+\dd{t}$ as $\rho_1(t) \dd{t}$. Since the transition of the memory from the collective excited state $\ket*{e_\chi}$ to the fundamental state $\ket{G}$, with the corresponding emission of a photon in the detection mode, is irreversible, from Eq.~\eqref{phi} we conclude that the probability that the photon has been emitted at some instant before time $t$ is given by $\int \dd[3]{\vb{k}} \abs{\gamma(t;\vb{k}) \Phi(\vb{k})}^2$.
So, by applying Weisskopf--Wigner method approximations, we get
\begin{equation}
	\label{probdensity}
	\rho_1(t) = \Gamma_\chi \abs{\beta(t)}^2,
\end{equation}
where
\begin{equation}
	\label{decayrate}
 	\Gamma_\chi = 2\pi \int \dd[3]{\vb{k}} \delta(\omega - \omega_{ge}) \abs{g(\vb{k}) \Phi(\vb{k})}^2.
\end{equation}

If we have a single atom, then only one of the coefficients $\beta_i$ is $1$ and all others are null, such that in this case we have $\abs{\Phi(\vb{k})}^2=1$ and Eq.~\eqref{decayrate} reduces to
\begin{equation}
	\Gamma = 2\pi \int \dd[3]{\vb{k}} \delta(\omega  - \omega_{ge}) \abs{g(\vb{k})}^2,
\end{equation}
which is the decay rate of an isolated atom \cite{scully1997quantum}.
In the scalar field approximation, the $g(\vb{k})$ depends only on the modulus of $\vb{k}$, which is fixed by the delta function $\delta(\omega  - \omega_{ge})$ at the resonance frequency. Then, the integration over all wavevector directions in Eq.~\eqref{decayrate} is affected only by $\abs{\Phi(\vb{k})}^2$ such that we can rewrite Eq.~\eqref{decayrate} as $\Gamma_{\chi} = \chi\Gamma$, where
\begin{equation}
	\label{fatorsuper}
	\chi = \frac{1}{4\pi k^2_{ge}} \int_{\abs{\vb{k}} = k_{ge}} \dd[2]\vb{k} \abs{\Phi(\vb{k})}^2,
\end{equation}
$k_{ge}$ being the wavenumber of the emitted photon. The parameter $\chi$ quantifies the system superradiance, such that we have a superradiant photon emission whenever $\chi > 1$.

From the calculations in the Appendix, we can write the ``superradiance enhancement'' as a function of the waist of the Gaussian mode of photon $1$, $w_0$, and of the effective number of atoms that interact with this mode: $N$
	\begin{equation}
	\label{chi}
	\chi \approx 1 + \frac{N}{2w_0^2k^2_{ge}}.
\end{equation}
The value for $\chi - 1$ obtained here is half the value used in our previous works \cite{1367-2630-15-7-075030, PhysRevA.90.023848, PhysRevLett.120.083603}. As the method for computing $\chi$ here is different from the method used in Ref.~\cite{1367-2630-15-7-075030}, with a different set approximations, this difference is not a surprise. Since the formalism used here is much cleaner, we are more confident in the expression above for the superradiance enhancement $\chi$ in relation to the previous expression used in our works. 

Now we apply the Weisskopf--Wigner method to find the $\beta(t)$ in Eq.~\eqref{probdensity}, resulting in a differential equation for $\beta(t)$ only in terms of the amplitudes describing the atomic cloud: 
\begin{equation}
	\label{equations1}
	\dot{\beta}(t) = -\frac{\Omega_0}{2} \alpha(t) - \frac{\chi\Gamma}{2} \beta(t).
\end{equation}
By defining the effective Rabi frequency $\Omega$ and a phase constant $\phi$ as
\begin{equation}
	\label{omega}
	\Omega^2 = \Omega_0^2 - \pqty{\frac{\chi \Gamma}{2}}^2, \qquad \sin\phi = \frac{\chi\Gamma}{2\Omega_0},
\end{equation}
and using the initial conditions $\alpha(0) = 1$ and $\beta(0) = 0$, we solve the system of differential Eqs.~\eqref{newequationsa} and \eqref{equations1}:
\begin{subequations}
	\label{solutions1}
	\begin{align}
		\alpha(t) & = \sec\phi\, \me^{-\frac{\chi\Gamma}{4} t} \cos\pqty{\frac{\Omega}{2}t - \phi}, \\
		\beta(t) & = \sec\phi\, \me^{-\frac{\chi\Gamma}{4} t} \sin\pqty{\frac{\Omega}{2}t}.
	\end{align}
\end{subequations}
Substituting the above result for $\beta(t)$ in Eq.~\eqref{probdensity} we obtain
\begin{equation}
	\label{p1}
	\rho_1(t) = a \, \me^{-\frac{\chi\Gamma}{2} t} \sin^2\pqty{\frac{\Omega}{2} t},
\end{equation}
where $a = \sec^2\phi = \chi\Gamma\Omega_0^2/\Omega^2$ is a normalization constant.

Here is a simple physical explanation for this expression. The read laser beam induces Rabi oscillations on the atom initially left at level $\ket{s}$ on the writing stage, no matter which atom is this. When the atom is in the excited  level $\ket{e}$, it can decay to the ground level $\ket{g}$ with the superradiant rate $\chi\Gamma$.  The superradiant factor $\chi$ results from the coherent sum of the probability amplitudes of emission by each atom that participates of the collective atomic state. A constructive interference of these possibilities increases the decay rate. The probability density of emission is then simply given by the probability that an atom is found on the excited state $\ket{e}$, given by $\sin^2\pqty{\Omega t/2}$, times the exponential superradiant decay $\me^{-\chi\Gamma t/2}$. The prediction of Eq.~\eqref{p1}, that was deduced in Ref.~\cite{1367-2630-15-7-075030} in a more complicated way, was successfully tested in previous works from our group \cite{1367-2630-15-7-075030,PhysRevA.90.023848,PhysRevLett.120.083603}. We will compare this prediction with other experimental results in Sec.~\ref{sec:experiments}.

\section{Theory for Two-Photon Superradiance}\label{theory2}

\subsection{Writing in the memory}

During the write process of the memory, it may happen that the atomic cloud scatters more then one photon into mode 1. Since the cloud has a huge number of atoms, we can approximate that each photon scattered to mode 1 leaves one excitation in the atomic mode $a$ defined in Eq.~\eqref{1a}. Disregarding other light and atomic modes that can be excited in the process, since these will not be relevant for the experiments we want to model, the system state at this stage can be written as 
\begin{equation}
	\label{e1}
	\ket{\Psi_{a,1}} = \sqrt{1 - p} \sum_{n=0}^{\infty} p^{n/2} \ket{n_a , n_1},
\end{equation}
in a superposition state of having with $n$ excitations stored in the collective mode $a$ and $n$ photons emitted in mode $1$ with different values of $n$. For $p\ll 1$, $p$ represents the probability of having a single excitation both in the atomic mode $a$ and in the photon mode $1$. Since the detectors do not distinguish the number of photons detected, a single photon detection in mode 1 ideally projects the ensemble in the state
\begin{equation}
	\label{e2}
	\ket{\psi_1} \propto \ket{1_a} + p^{1/2} \ket{2_a} + p \ket{3_a} + \cdots \, .
\end{equation} 
The treatment of the previous section is valid when $p\ll1$ and we have a state close to $\ket{1_a}$ from  Eq.~\eqref{1a}.

Two detections in mode 1 during the write process, on the other hand, projects the atomic ensemble in the state
\begin{equation}
	\label{e3}
	\ket{\psi_2} \propto \ket{2_a} + p^{1/2} \ket{3_a} + \cdots \, .
\end{equation} 
When $p\ll1$ we have a state close to $\ket*{2_a}$, that can be written as
\begin{equation}
	\label{ss}
	\ket{2_a} \approx \ket*{s_\chi s_\chi} \equiv \frac{1}{\sqrt{2}} \sum_{i,j} \alpha_i \alpha_j \, \ket*{s_i s_j},
\end{equation}
with $\alpha_i$ given by Eq.~\eqref{alphai} and $\ket*{s_i s_j}$ representing a state with atoms $i$ and $j$ on level $\ket{s}$ and all others on level $\ket{g}$. The factor $1/\sqrt{2}$ comes from the normalization condition $\braket*{s_is_j}{s_{i'}s_{j'}} = \delta_{ii'}\delta_{jj'} + \delta_{ij'}\delta_{ji'}$  (with $i \ne j$ and $i' \ne j'$). Note that this state is not exactly realizable, since we should have $i \neq j$ in Eq.~\eqref{ss}. But when the number of atoms is huge, as it is in the experiments we want to model, there is no significant difference.

\subsection{Reading the memory}

When the read process begins, the ensemble evolves in the interaction picture accordingly to the interaction Hamiltonian from Eq.~\eqref{hint}. The first term keeps the state in a subspace where the number of photons is definite, but the atoms may transit between the auxiliary $\ket{s}$ and excited $\ket{e}$ levels. The second term causes the transition between the excited level and the ground level $\ket{g}$ of each atom with a corresponding photon emission, moving the state between subspaces counting different numbers of photons. This analysis is valid already in the single-photon case, but it is more useful here given the increased complexity of the problem. 
	
If the system initial state is given by Eq.~\eqref{ss}, the evolution by the Hamiltonian from Eq.~\eqref{hint} leads the system to the general state
\begin{equation}
	\label{psi2}
	\ket{\psi_2(t)} = \ket*{\psi_{2a}(t)} + \ket*{\psi_{1a,1f}(t)} + \ket*{\psi_{2f}(t)},
\end{equation}
where $\ket*{\psi_{2a}(t)}$ corresponds to a state with zero photons and two excitations in the memory, $\ket*{\psi_{1a,1f}(t)}$ a state with one photon and one excitation in the memory and $\ket*{\psi_{2f}(t)}$ a state with two photons and no excitation in the memory. Using the same approximations that we have made in the previous section, considering the evolution of collective atomic states, we can write
\begin{equation}
	\label{psi20}
	\ket{\psi_{2a}(t)} = \lambda(t) \ket*{s_\chi s_\chi} + \mu(t) \ket*{s_\chi e_\chi} + \nu(t) \ket*{e_\chi e_\chi}
\end{equation}
with $\ket*{s_\chi s_\chi}$ given by Eq.~\eqref{ss},
\begin{equation}
	\ket*{s_\chi e_\chi} \equiv \sum_{ij} \alpha_i \beta_j \ket*{s_i e_j} \qand \ket*{e_\chi e_\chi} \equiv  \frac{1}{\sqrt{2}} \sum_{ij} \beta_i \beta_j \ket*{e_i e_j},
\end{equation}
with $\beta_i$ given by Eq.~\eqref{betai} and an obvious generalization of the notation used in Sec. \ref{theory1}. We can also write 
\begin{equation}
	\label{psi21}
	\ket*{\psi_{1a,1f}(t)} = \int \dd[3]{\vb{k}} \bqty{\xi(t;\vb{k}) \Phi(\vb{k}) \ket*{s_\chi,\vb{k}} + \zeta(t;\vb{k}) \Phi(\vb{k}) \ket*{e_\chi,\vb{k}}},
\end{equation}
and
\begin{equation}
	\label{psi22}
	\ket*{\psi_{2f}(t)} = \frac{1}{\sqrt{2}} \iint \dd[3]{\vb{k}} \dd[3]{\vb{k}'} \eta(t;\vb{k},\vb{k}') \Phi(\vb{k}) \Phi(\vb{k}') \ket*{g,\vb{k}\vb{k}'},
\end{equation}
with $\Phi(\vb{k})$ given by Eq. \eqref{phi} and an obvious generalization of the notation used in Sec. \ref{theory1}.

To find the temporal wavepacket of the emitted photons, we will consider the system dynamics using the quantum trajectories method \cite{0954-8998-6-1-003, PhysRevLett.68.580, PhysRevA.47.449}. The reason is that the photon detectors continuously monitor the system, such that while no photon was detected, we can assume that the system is in a subspace with zero photons and two excitations in the memory or one photon (that could have been emitted in an infinitesimal instant before) and one excitation in the memory. After the first photon is detected, the system collapses in a subspace with one excitation in the memory and zero photons, or zero excitations in the memory and a new photon in the field. In this way, the first photon emission and the second photon emission are treated separately. It is implicit in the treatment that we consider a very low probability that both photons are emitted at the same time.

\subsubsection{First emission}

As previously stated, to find the probability density of detection as a function of time of the first emitted photon, we use the Schr\"odinger equation with the Hamiltonian of Eq.~\eqref{hint} acting on the state of Eq.~\eqref{psi2} projected on the subspace that excludes the portion $\ket*{\psi_{2f}t)}$. We then arrive at the following differential equations for the coefficients present in Eqs.~\eqref{psi20} and \eqref{psi21}:
\begin{subequations}
	\begin{align}
		\label{newequations2a}
		\dot{\lambda(}t) & = \frac{\Omega_0}{\sqrt{2}} \mu(t), \\
		\label{newequations2b}
		\dot{\mu}(t) & = \frac{\Omega_0}{\sqrt{2}} \pqty{-\lambda(t) + \nu(t)} - \! \int \! \dd[3]{\vb{k}} g(\vb{k}) \abs{\Phi(\vb{k})}^2 \me^{-\mi (\omega  - \omega_{ge})t} \xi(t; \vb{k}), \\
		\label{newequations2c}
		\dot{\nu}(t) & = -\frac{\Omega_0}{\sqrt{2}} \mu(t) - \sqrt{2} \int \dd[3]{\vb{k}} g(\vb{k}) \abs{\Phi(\vb{k})}^2 \me^{-\mi (\omega  - \omega_{ge})t} \zeta(t; \vb{k}), \\
		\label{newequations2d}
		\dot{\xi}(t; \vb{k}) & = g^*(\vb{k})\me^{\mi (\omega  - \omega_{ge})t} \mu(t) + \frac{\Omega_0}{2} \zeta(t; \vb{k}), \\
		\label{newequations2e}
		\dot{\zeta}(t; \vb{k}) & = \sqrt{2} g^*(\vb{k})\me^{\mi (\omega  - \omega_{ge})t} \nu(t) - \frac{\Omega_0}{2} \xi(t; \vb{k}).
	\end{align}
\end{subequations}
Note that the second terms on the right hand side of Eqs.~\eqref{newequations2d} and \eqref{newequations2e} are the ones responsible for the Rabi oscillations between states $\ket*{s_\chi,\vb{k}}$ and $\ket*{e_\chi,\vb{k}}$. If we take into account that, with the use of the quantum trajectories method, we are treating the system before the first photon emission or just after its emission, these terms do not contribute to the system dynamics at this stage. In other words, we can say that, just before the photon emission, we have  $\zeta(t; \vb{k})= \xi(t; \vb{k})=0$, such that we can approximate Eqs.~\eqref{newequations2d} and \eqref{newequations2e} as
\begin{subequations}
	\begin{align}
		\label{xi}
		\dot{\xi}(t; \vb{k}) & = g^*(\vb{k})\me^{\mi (\omega  - \omega_{ge})t} \mu(t), \\
		\label{zeta}
		\dot{\zeta}(t; \vb{k}) & = \sqrt{2} g^*(\vb{k})\me^{\mi (\omega  - \omega_{ge})t} \nu(t).
	\end{align}
\end{subequations}

As in the single-photon case, the probability density of emission of the first photon can be written as
\begin{equation}
	\label{probdensityfirst}
	\rho_2(t) = \chi\Gamma \abs{\mu(t)}^2 + 2\chi\Gamma \abs{\nu(t)}^2,
\end{equation}
with $\chi$ given by Eq.~\eqref{fatorsuper}. Now we need to find the amplitudes $\mu(t)$ and $\nu(t)$. The Weisskopf--Wigner approximations are used again, then we get the following set of differential equations for the coefficients $\lambda(t)$, $\mu(t)$, and $\nu(t)$:
\begin{subequations}
	\label{equations2}
	\begin{align}
		\dot{\mu}(t) & = -\frac{\Omega_0}{\sqrt{2}} \lambda(t) + \frac{\Omega_0}{\sqrt{2}} \nu(t) - \frac{\chi\Gamma}{2} \mu(t), \\
		\dot{\nu}(t) & = -\frac{\Omega_0}{\sqrt{2}} \mu(t) - \chi\Gamma \nu(t),
	\end{align}
\end{subequations}
together with Eq.~\eqref{newequations2a}. From the initial conditions $\lambda(0) = 1$ and $\mu(0) = \nu(0) = 0$, we arrive at the following solutions as functions of $\alpha(t)$ and $\beta(t)$ from Eq.~\eqref{solutions1}
\begin{equation}
	\label{solutions2}
	\lambda(t) = \alpha^2(t), \quad \mu(t) = \sqrt{2} \alpha(t) \beta(t), \quad \nu(t) = \beta^2(t).
\end{equation}
This solution indicates that the dynamics of each excitation in the memory evolves independently of the other during the first emission.  

Using Eqs.~\eqref{solutions2} and \eqref{solutions1} in Eq.~\eqref{probdensityfirst}, we find that the probability density of the first photon detection at time $t_1$ is
\begin{equation}
    \label{p21}
	\rho^{(2)}_1(t_1) = a_1 \me^{-\chi\Gamma t_1} \sin^2 \pqty{\frac{\Omega t_1}{2}} \bqty{1 + b_1 \sin\pqty{\Omega t_1} + c_1 \cos\pqty{\Omega t_1}},
\end{equation}
$a_1 = 2 a=2\chi\Gamma\Omega_0^2/\Omega^2$, $b_1 = \chi\Gamma\Omega/2\Omega_0^2$, and $c_1 = -\chi^2 \Gamma^2/4\Omega_0^2$. By comparing the exponential decay on this expression with the exponential decay on Eq.~\eqref{p1}, we see that the decay rate for the first photon emission in the case of two excitations in the memory is twice the decay rate when we have only one excitation.
	
\subsubsection{Second emission}

With the detection of the first photon at time $t_1$, we can use the quantum trajectories method \cite{0954-8998-6-1-003, PhysRevLett.68.580, PhysRevA.47.449} to compute what is the system state just after this detection. During the detection, the system state is given by Eq.~\eqref{psi2} projected in the subspace with at most one photon in the system. The photon detection projects this state on a state with at least one photon in the system and removes this photon. So, according to Eq.~\eqref{psi21}, the system state after the first detection will be
\begin{equation}
	\ket*{\psi'_{1a}(t_1)} \propto \xi'(t_1) \ket*{s_\chi} + \zeta'(t_1) \ket*{e_\chi}.
\end{equation}
The coefficients $\xi'(t_1)$ and $\zeta'(t_1)$ can be computed from Eqs.~\eqref{newequations2d} and \eqref{newequations2e} considering that, in an infinitesimal time $\epsilon$ before the detection, only the coefficients $\lambda(t)$, $\mu(t)$, and $\nu(t)$ are different from zero. So we have $\xi'(t_1)/\epsilon\approx g^*(\vb{k})\mu(t_1 - \epsilon)$, $\zeta'(t_1)/\epsilon\approx\sqrt{2} g^*(\vb{k})\nu(t_1 - \epsilon)$, with a renormalization of the state. With the use of Eq.~\eqref{solutions2}, we obtain
\begin{equation}
	\label{psi21t1}
	\ket*{\psi'_{1a}(t_1)} \propto \alpha(t_1) \ket*{s_\chi} + \beta(t_1) \ket*{e_\chi},
\end{equation}
with $\alpha(t_1)$ and $\beta(t_1)$ given by Eqs.~\eqref{solutions1}. 

From time $t_1$ on, the system dynamics will be exactly as the one computed in Sec.~\ref{theory1}, when there was only one excitation stored in the memory, conditioned to the initial state \eqref{psi21t1}. Comparing to Eq.~\eqref{psi1}, we see that this initial state is the same as in the single excitation case treated in Sec.~\ref{theory1} where there was no detection until time $t_1$. So the probability density of detection of the second photon at a time $\tau$ after the detection of the first, conditioned to the detection of the first at time $t_1$, is given by
\begin{equation}
	\label{p22}
	\rho_2(t_1+\tau | t_1) = \frac{\rho_1(t_1+\tau)}{\int_{0}^\infty \rho_1(t_1+\tau) \dd{\tau}},
\end{equation}
where the integral in the denominator is a normalization factor. Using Eq.~\eqref{p1}, it can be shown that 
\begin{equation}
	\label{norm}
	\int_{0}^\infty \rho_1(t_1+\tau) \dd{\tau} = \abs{\alpha(t_1)}^2 + \abs{\beta(t_1)}^2,
\end{equation}
with $\alpha(t_1)$ and $\beta(t_1)$ given by Eqs.~\eqref{solutions1}. Since the probability density of detecting the first photon at time $t_1$ and the second one at time $t_2 = t_1 + \tau$, given $\tau > 0$, is given by $\rho^{(2)}(t_1, t_1 + \tau) = \rho^{(2)}_2(t_1 + \tau | t_1) \rho^{(2)}_1(t_1)$. Using Eqs.~\eqref{p21}, \eqref{p22}, and \eqref{norm}, we find
\begin{equation}
	\label{probconj}
	\rho^{(2)}(t_1,t_1 + \tau) = 2 \rho^{(1)}(t_1) \rho^{(1)}(t_1 + \tau),
\end{equation}
written as function of $\rho^{(1)}(t)$ from Eq.~\eqref{p1}. If we write the probability density of photon detections at times $t_a$ and $t_b$, no matter if $t_a<t_b$ or $t_b<t_a$, the above probability density is 
\begin{equation}
	\label{independent}
	\rho^{(2)}(t_a,t_b) = \rho^{(1)}(t_a) \rho^{(1)}(t_b),
\end{equation}
with no correlation between the two photon emissions. We conclude that both photons are emitted independently and in the same spatiotemporal mode, with superradiant features. The quantum memory acts as a superradiant source of a Fock state of light with two photons in this case, what can have significant applications in the search for reliable Fock-state sources.

Another quantity that we can deduce is the probability of detection of the second photon an interval $\tau$ after the first, no matter at what time the first detection occurs:
\begin{equation}\
	\rho_2^{(2)}(\tau) = 2\int_0^\infty \rho^{(1)}(t_1) \rho^{(1)}(t_1 + \tau) \dd{t}_1.
\end{equation}
It can be shown that the integral results in
\begin{equation}
	\label{second}
	\rho_2^{(2)}(\tau) = a_2 \me^{-\chi\Gamma \tau/2} \bqty{ 1 + b_2 \sin\pqty{\Omega\tau} + c_2 \cos\pqty{\Omega \tau}}
\end{equation}
with $a_2 = a \Omega_0^2/2(\Omega^2 + \chi^2\Gamma^2)$, $b_2 = 3\chi\Gamma \Omega/4\Omega_0^2$, and $c_2 = 3\Omega^2/2\Omega_0^2 - 1$.

\section{Experiments}
	\label{sec:experiments}

In the following, we describe our series of experiments. Single- and two-photon superradiances are investigated simultaneously through the same experiment, just being heralded by different types of events during the writing process. The theory in Secs.~\ref{theory1} and \ref{theory2} is more complicated than the corresponding one on Ref.~\cite{1367-2630-15-7-075030} and derives single- and two-photon wavepackets together from the same set of parameters, with a more significant number of experimental curves being adjusted by the same number of parameters as in our previous studies on single-photon superradiance \cite{1367-2630-15-7-075030,PhysRevA.90.023848}. The theory in Secs.~\ref{theory1} and ~\ref{theory2} is derived neglecting any effect of reabsorption of photons by the ensemble in the reading process, in which case no correlation between the two photons should be observed. In this case, the two-photon wavepacket should be merely the product of two independent single-photon emissions, as stated in Eq.~\eqref{independent}. In this way, we can investigate possible correlations between the two photons emitted in the reading process by comparing the experimentally measured two-photon wavepackets with the expected curves coming from the experimentally measured single-photon wavepackets and assuming independence between the emitted photons.

The experiments were conducted in two stages. In the first one, described in Sec.~\ref{linear}, we employed an experimental configuration most similar to Ref.~\cite{1367-2630-15-7-075030}, i.e., linear polarizations on all fields and without optical pumping to a specific Zeeman sublevel. Differently from our previous study, on the other hand, our ensemble now consists of a cloud of cold rubidium 87 atoms, and we did not employ any frequency filter on field $1$. The removal of such filter resulted in a significant improvement on the count rates for the two-photon wavepacket, effectively making the experiment feasible. The results on Sec.~\ref{linear} allowed for an investigation on the independence of the photons on the two-photon superradiant emission, but not for a direct comparison with the final theoretical expressions deduced on Sec.~\ref{theory2}. The experimental wavepackets presented beatings coming from the various Rabi oscillations involved in the reading process through different Zeeman sublevels, while the theory considered a simpler three-level atom.

In the second stage described in Sec.~\ref{circular}, we then optically pumped the atoms to a specific Zeeman sublevel and employed circular polarizations for all fields, in order to guarantee a single Rabi frequency on the reading process. We also observed a significant enhancement on the Rabi oscillations by turning on faster the read beam, through the use of an in-fiber intensity modulator, which approximates the experiment to the theoretical assumption of a step function for the turning on of the read field. With these modifications, we were finally able to directly compare our experimental wavepackets to the theory of Sec.~\ref{theory2}, reaching a reasonable agreement. As in Sec.~\ref{linear}, we also analyzed for this new configuration the independence of the two emitted photons in the two-photon superradiant wavepacket, by directly comparing it to the expected results obtained from the experimental single-photon wavepackets.

\subsection{First configuration: linear polarizations}
\label{linear}

In our first series of experiments we employed a cold ensemble of rubidium 87 atoms, obtained from a Magneto-Optical Trap (MOT). The trap was kept on for $23$~ms, before the trapping beams and the magnetic field were turned off for $2$~ms. The MOT repumping light, resonant with the $\ket*{5S_{1/2}(F=1)} \rightarrow \ket*{5P_{3/2}(F' = 2)}$ transition, is kept on for an extra $0.9$~ms to pump all atoms to the $\ket*{5S_{1/2}(F=2)}$ state, the initial $\ket*{g}$ state of our scheme. After the repumping light is turned off, the ensemble is kept in the dark from all MOT light fields during $1.1$~ms, the period in which the experiment takes place. The residual magnetic field is minimized up to a ground state linewidth of about $100$~kHz, corresponding to magnetic fields on the order of $36$~mG, by means of three pairs of compensating coils in Helmholtz configuration and performing microwave spectroscopy between the two hyperfine ground states~\cite{PhysRevA.94.063834}. Such linewidth should lead to a coherence time on the order of $1.6$~$\mu$s, much larger than the separation between write and read processes employed below.

\begin{figure}[ht]
	\centering 
	\includegraphics[width=8cm]{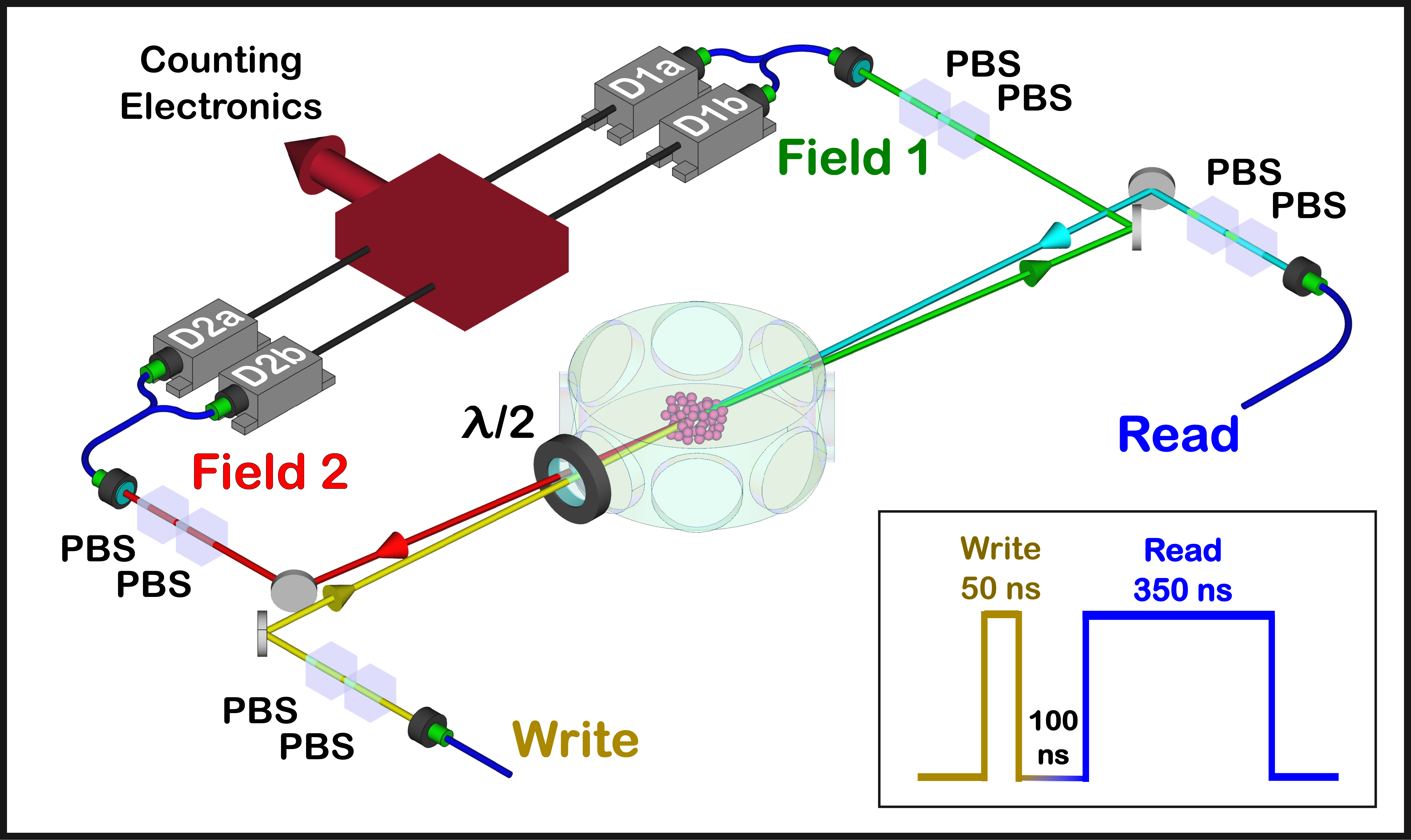}
	\caption{
		\label{setup1}
		Experimental setup for the first configuration considered in our experiments: linear polarizations. PBS stands for polarizing beam splitter, and $\lambda /2$ for half-wave plate.}
\end{figure}

The general setup for the experiment is shown in Fig.~\ref{setup1}. During the dark period of the MOT, a sequence of 1000 write and read pulses excite the ensemble. The write pulse has a duration of about $50$~ns and is $40$~MHz red detuned to the $\ket*{g} \rightarrow \ket*{e}$ transition, with $\ket*{e} = \ket*{5P_{3/2}(F' = 2)}$. Its duration and frequency are controlled by a sequence of two Acoustic Optical Modulators (AOM), one of them in a double-pass configuration. The $4\sigma$ diameter of the write beam in the MOT region is $420$~$\mu$s. As a result of its action, photons may be emitted in the mode of field $1$ coupled to an input of a single-mode fiber beam splitter that divides the output to two Avalanche PhotoDetectors (APD), D$_{1a}$ and D$_{1b}$. The field responsible for our signal is emitted in the $\ket*{e} \rightarrow \ket*{s}$ transition, with $\ket*{s} = \ket*{5S_{1/2}(F=1)}$.

The Optical Depth ($OD$) of the atomic ensemble is measured by sending a $1$~$\mu$s pulse in the write field spatial mode and observing its absorption after propagation in the atomic ensemble. This parameter is crucial~\cite{PhysRevA.90.023848} since it is proportional to the number of atoms in the system and is easily accessible experimentally. The number of atoms control, for example, the decay time of the collective superradiant emission~\cite{1367-2630-15-7-075030}. In order to characterize the optical depth in our experiment we followed the approach of Ref.~\cite{1402-4896-81-2-025301}. For the experiments on this subsection, we obtained $OD = 31$.

In the MOT region, field $1$ has a $4\sigma$ diameter of $150$~$\mu$m and pass through the middle of the write beam forming an angle of $3^{\circ}$ with it. The polarizations of write beam and field $1$ are linear and orthogonal to each other, with extinction ratios on the order of $10^5$ for the orthogonal polarizations in each field. The transmission through a pair of Polarizing Beam Splitters (PBS) achieves this degree of polarization, and a half-wave plate ($\lambda/2$) does the rotation of the polarization of the write beam right before the vacuum chamber and a quarter-wave plate ($\lambda/4$, not shown on Fig.~\ref{setup1}) to correct for small polarization distortions on the optical pathway. 

\begin{figure}[ht]
	\centering 
	\includegraphics[width=7.0cm]{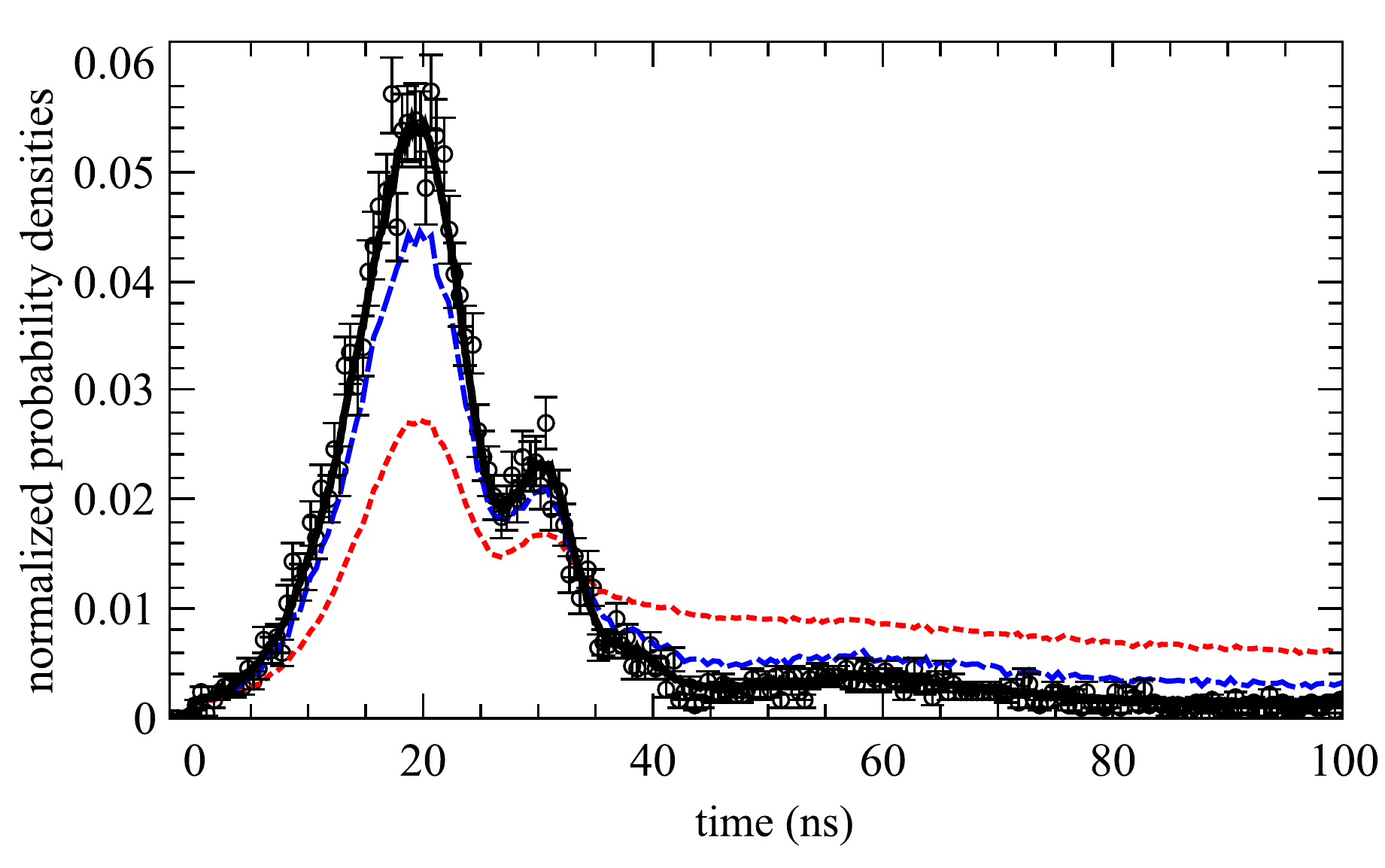}
	\caption{
		\label{linear01}
		Single-photon wavepacket [$\rho^{(1,1)}(t)$, continuous black curve] of field $2$ compared with three different two-photon wavepackets on field $2$: one not heralded on any event on field $1$ (dashed red curve), the other heralded by just one event on field $1$ (traced blue curve), and the last one heralded on two events on field $1$ (black circles). The two detection events on the two-photon wavepackets were plotted independent of each other, corresponding then to the normalized probability densities $\rho^{(2,0)}(t)$, $ \rho^{(2,1)}(t)$ and $\rho^{(2,2)}(t)$. Linear polarizations were applied for write and read fields.}
\end{figure}

After the write field is turned off by about $100$~ns, the read pulse is turned on for $350$~ns (see inset of Fig.~\ref{setup1}) by a single AOM. It is resonant to the $\ket*{s} \rightarrow \ket*{e}$ transition and maps the collective state with one excitation in $\ket*{s}$ to a collective state with one excitation in $\ket*{e}$, which then decays superradiantly back to the initial state $\ket*{g}$~\cite{PhysRevA.90.023848}. We call field $2$ this final superradiant emission of the overall, parametric four-wave-mixing process~\cite{PhysRevA.72.053809}. In the case of a single heralding event on field $1$, field $2$ results in the single-photon wavepacket $\rho^{(1,1)}$~\cite{PhysRevA.90.023848} plotted as the continuous black curve on Fig.~\ref{linear01}. The overall probability to detect an event on field $1$ is here $p_1 = 0.016$, and the second-order auto-correlation for the conditioned field $2$ is $g_2^c = 0.487 \pm 0.004$. The value $g_2^c < 1$ indicates the sub-poissonian character of field $2$ as it enters the single-photon regime. The total conditional probability to detect a photon on field $2$, once a photon was previously detected on field $1$, is $P_c = 9.5\%$. The rate of coincidence between single events on fields $1$ and $2$ is $60$~Hz.

We do not attempt to match this single-photon wavepacket to the theory of Ref.~\cite{PhysRevA.90.023848}, presented also on Sec.~\ref{theory1}, since we clearly observe beatings between different Rabi frequencies. This indicates that our previous approximation of a simple three-level system is not valid in this case. In order to compare to the two-photon wavepackets, $\rho^{(1,1)}$ provides the probability density at time $t$, counted from the beginning of the read pulse, to measure a single photon in field $2$ once a single photon was detected in field $1$, normalized by the area under the curve, which equals $P_c$.
  
For the two-photon wavepackets, we plot three different wavepackets. The dashed red curve gives a wavepacket of two-photon events on field $2$ not conditioned on any event on field $1$ [$\rho^{(2,0)}(t)$]. The traced blue curve gives the same quantity heralded by a single event on field $1$ [$\rho^{(2,1)}(t)$], and the black circles the same quantity conditioned on two events on field $1$ [$\rho^{(2,2)}(t)$]. We plot in Fig.~\ref{linear01} the results for the two detections on field $2$ independent one from the other. To be more explicit, note that the complete two-photon wavepacket has two characteristic times $t_a,t_b$ corresponding to the instants of each of the two detections. For the wavepacket conditioned on two detections on field $1$, if ${\rho}^{(2,2)}(t_a,t_b)$ is the normalized probability density giving the complete two-photon wavepacket, then the quantity plotted on Fig.~\ref{linear01} is
\begin{equation}
	\label{rho22}
	\rho^{(2,2)}(t) = \frac{1}{2}\bqty{\int_0^{\infty} \rho^{(2,2)}(t,t_a)\,\dd{t}_a + \int_0^{\infty} \rho^{(2,2)}(t_b,t) \dd{t}_b} .
\end{equation}
The other two-photon wavepackets on Fig.~\ref{linear01} are similarly defined. For the wavepacket given by the black circles on Fig.~\ref{linear01} we had $2541$ four-photon coincidences, generated at a rate of $30$~mHz.

From Fig.~\ref{linear01} we then notice that only ${\rho}^{(2,2)}$ has a direct correspondence to the single-photon wavepacket, as expected from the relation
\begin{equation}
	\label{ind}
	\rho^{(2,2)}(t_a,t_b) = \rho^{(1,1)}(t_a)\; \rho^{(1,1)}(t_b) , 
\end{equation}
that is equivalent to Eq.~\eqref{independent}, resulting, via Eq.~\eqref{rho22}, in
\begin{equation}
	\rho^{(2,2)}(t) = \rho^{(1,1)}(t) .
\end{equation}
Figure~\ref{linear01} provides then the first experimental evidence for the validity of Eq.~\eqref{independent}, i.e., that the two photons in the two-photon wavepacket are emitted independently of each other at the same temporal mode. It also explicitly demonstrates the strong effect the appropriate heralding has over the state of field $2$.

\begin{figure}[ht]
	\centering 
	\includegraphics[width=7.0cm]{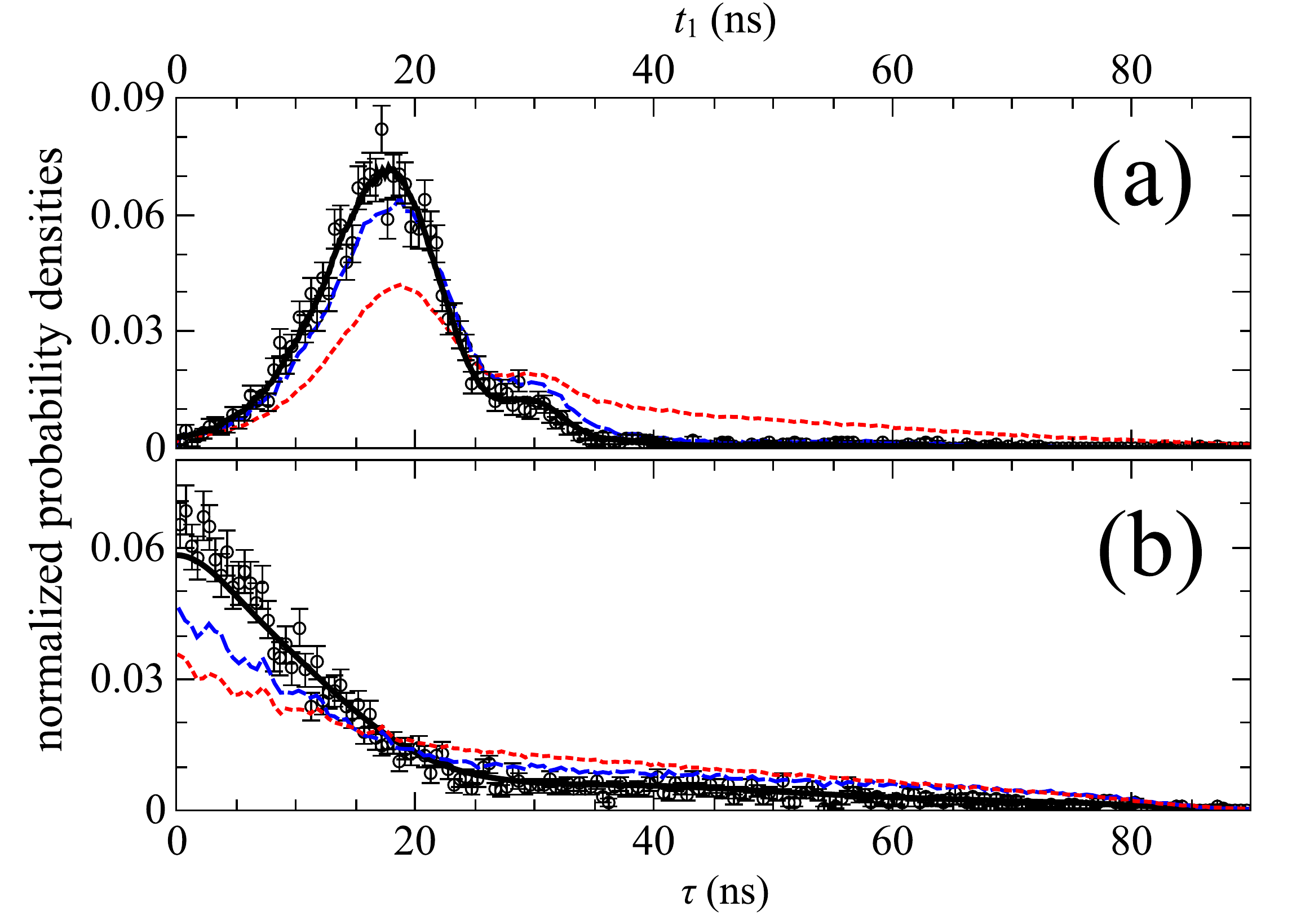}
	\caption{
		\label{linear02}
		Two-photon wavepackets for the configuration of linear polarizations. (a) Normalized probability densities $\rho^{(2,i)}_{1} (t_1)$ to detect the first field-$2$ photon at time $t_1$, conditioned on the detection of zero (dashed red curve), one (traced blue curve) or two (black circles) photons in field 1. (b) Normalized probability densities $ \rho^{(2,i)}_{2} (\tau)$ to detect the second field-$2$ photon at a time $\tau$ after the first detection, conditioned on the detection of zero (dashed red curve), one (traced blue curve) or two (black circles) photons in field 1. The continuous black curves in (a) and (b) are obtained from Eq.~\eqref{ind} and the measured single-photon wavepacket (continuous black curve in Fig.~\ref{linear01}.)}
\end{figure}

The complete wavepacket for the situation of Fig.~\ref{linear01} is plotted in Fig.~\ref{linear02}. The information on the two detections in the pair is divided in two parts. We plot in Fig.~\ref{linear02}(a) the probability distributions $\rho^{(2,i)}_1(t_1)$ to detect the first photon at $t_1$, and in Fig.~\ref{linear02}(b) the probability distributions $\rho^{(2,i)}_2(\tau)$ to detect the second photon a time $\tau$ after the first detection, for the wavepacket conditioned on $i$ photons detected on field $1$ during the write process. The black curves are the respective probability distributions obtained numerically from Eq.~\eqref{ind}, with $\rho^{(1,1)}(t)$ given by the continuous black curve of Fig.~\ref{linear01}.

As observed previously on Fig.~\ref{linear01}, only the proper heralding with two events on field $1$ leads to a two-photon wavepacket corresponding to two independent emissions on field $2$, as represented by the black curves. The crucial curve is the one for $\rho^{(2,2)}_2(\tau)$, since we would expect any deviation due to interaction between the outgoing photons, mediated by the ensemble, as being more pronounced when they are emitted close together around $\tau=0$. It is noticeable that the curves depending on $\tau$ present the largest variations with the number of heralding events. The data for $\rho^{(2,2)}_2(\tau)$ around $\tau = 0$ also seems to deviate a little bit for larger values from the black curve assuming independence between the two field-$2$ photons. However, the deviations are of the order of the error bars. Largely, our experimental data corroborate the independence of the emission for the two outgoing photons in the reading process, after preparation of collective symmetrical state with two excitations.

\subsection{Second configuration: Circular polarizations with optical pumping}
	\label{circular}

\begin{figure}[ht]
	\centering 
	\includegraphics[width=8cm]{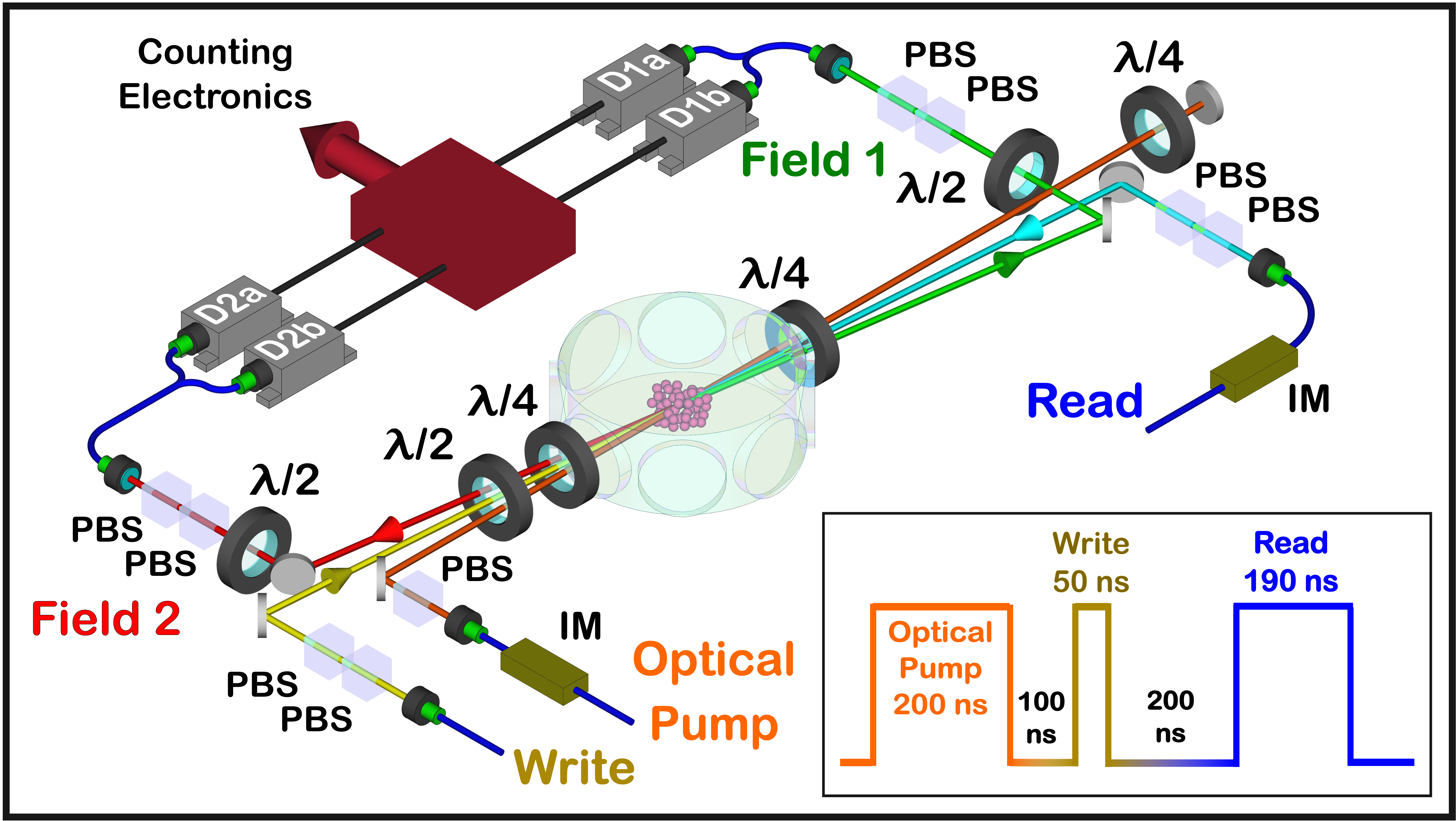}
	\caption{
		\label{setup2}
		Experimental setup for the second configuration considered in our experiments, with circular polarizations and optical pumping. PBS stands for polarizing beam splitter, $\lambda/2$ for half-wave plate, $\lambda/4$ for quarter-wave plate, and IM for intensity modulator.}
\end{figure}

The configuration on Sec.~\ref{linear} was interesting because it allowed us to address the question of independence of emissions of the two photons in the pair independently of any detailed theory for the overall process. It was also our configuration of maximum efficiency, for which we had highest rates of four-photon coincidences. However, in order to address directly the superradiant aspect of the process, our approach introduced on Refs.~\cite{1367-2630-15-7-075030} and~\cite{PhysRevA.90.023848} involves necessarily a comparison to a theoretical expression to extract the value of the superradiance enhancement $\chi$. In this way, we had to prepare the ensemble and modify the excitation fields so as to have a simple three-level atom participating in the whole process. This was done by first introducing an optical pumping field to leave the atoms at the state $\ket*{g} = \ket*{5S_{1/2}(F = 2, m_F = -2)}$, see Fig.~\ref{setup2}. The optical pumping is circularly polarized and red detuned to the $F=2 \rightarrow F' = 3$ transition, to reduce its mechanical action over the atoms. It consists of a $200$~ns pulse acting on the ensemble right before the write field (inset of Fig.~\ref{setup2}). 

The circularly polarized write field acts on the transition $\ket*{g} \rightarrow \ket*{e}$, with $\ket*{e} = \ket*{5P_{3/2}(F' = 2, m_F = -1)}$, for $50$~ns. We detect field $1$ during the write pulse with $\sigma^-$ polarization, i.e., in the transition $\ket*{e} \rightarrow \ket*{s}$ with $\ket*{s} = \ket*{5S_{1/2}(F=1, m_F=0)}$. Right after the write pulse, a $\sigma^-$ read pulse acts on the ensemble for $190$~ns exciting the transition $\ket{s} \rightarrow \ket{e}$. As a result, field $2$ is emitted with $\sigma^+$ polarization on the transition $\ket*{e} \rightarrow \ket*{g}$. Fast Intensity Modulators (IM) were introduced on the optical pathways of both optical pumping and read fields to ensure a faster turning on and off, and to decrease any residual background coming from these fields. For the read field, we particularly found that the IM introduction enhanced the visibility of the Rabi oscillations, which we attribute to a better approximation of the excitation field to a step function. For maximum optical depth of $OD_1 = 31.4 \pm 0.4$, we now have a total conditional probability of $P_c = 6.3\%$, a two-photon coincidence rate of $40$~Hz, and a four-photon coincidence rate of $14$~mHz. The optical depth was measured on the transition $\ket*{g} \rightarrow \ket*{5P_{3/2}(F = 3, m_F = -3}$, and the number of atoms in the modes of fields $1$ and $2$ was estimated as $N \approx 1.9 \times 10^6$ for this $OD$~\cite{PhysRevA.90.023848}. The read power was $P_R = 3.9$~mW. The sub-poissonian character of field $2$ was verified by measuring $g_2^c = 0.405 \pm 0.004 < 1$ for this situation. A lower optical depth, $OD_2 = 15.9 \pm 0.5$, was also employed for comparison. $OD$ was tuned by changing the power of the trapping light of the magneto-optical trap~\cite{PhysRevA.90.023848}. All other parameters were kept constant for the curves with $OD_1$ and $OD_2$.

\begin{figure}[ht]
	\centering 
	\includegraphics[width=7.0cm]{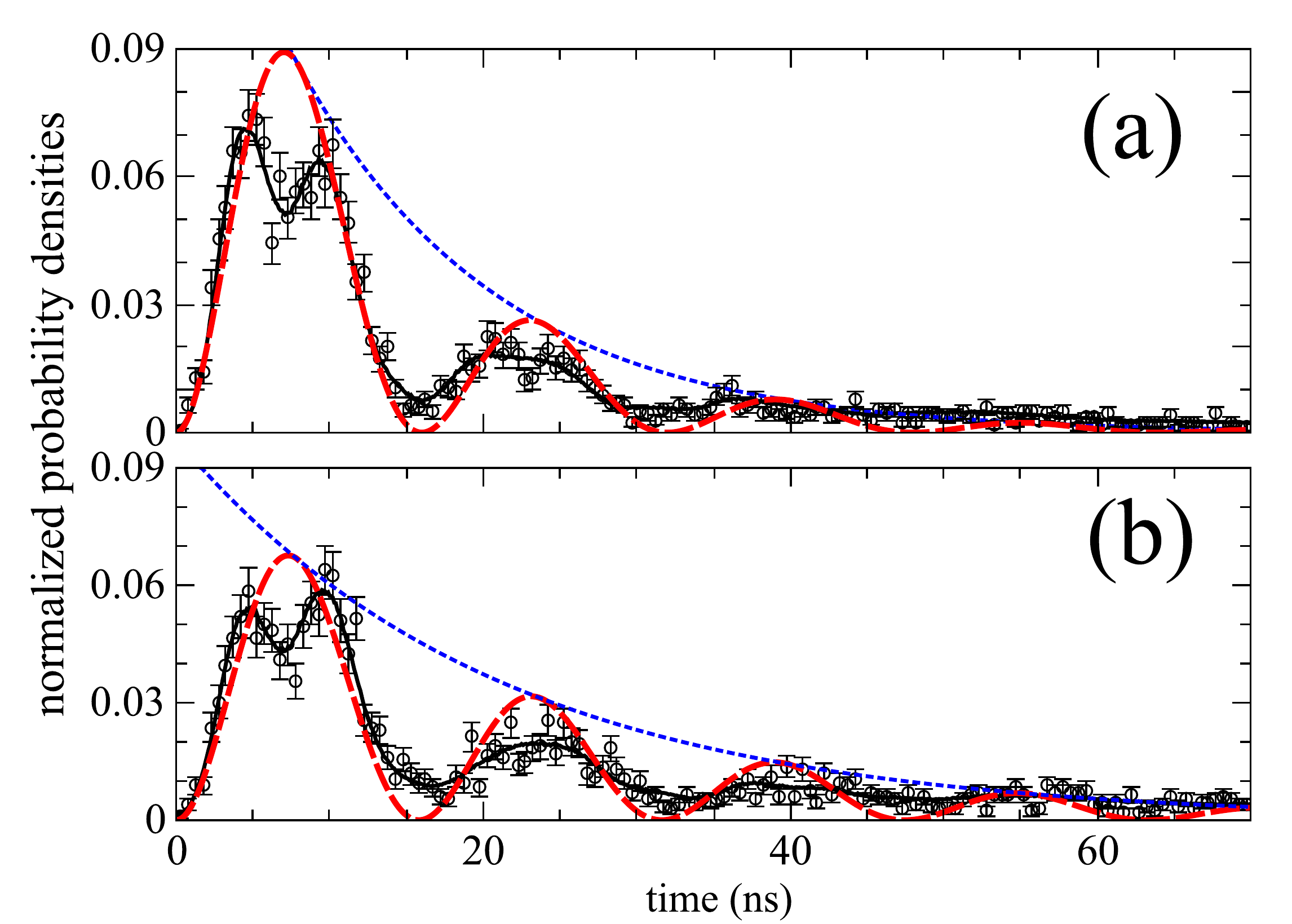}
	\caption{
		\label{circ01}
		Single-photon wavepackets (continuous black curves curves) for $OD_1$ (a) and $OD_2$ (b), compared to their respective independent-detections two-photon wavepackets $\rho^{(2,2)}(t)$ (black circles). Traced red lines provide the corresponding theoretical curves according to Eq.~\eqref{p1}. Dashed blue lines plot the respective pure exponential decays.}
\end{figure}

For this new configuration, Fig.~\ref{circ01} plots then ${\rho}^{(2,2)}(t)$ together with $\rho^{(1,1)}(t)$ for the two optical depths. As in Fig.~\ref{linear01}, the close similarity between the single- and two-photon wavepackets for both optical depths reinforce the independence between the two emitted photons. Now, however, we clearly have a distinct Rabi oscillation and the corresponding decay envelope, which can be compared to the theory of Sec.~\ref{theory1} given by the traced red line of Fig.~\ref{circ01}. The theory here corresponds to Eq.~\eqref{p1} for the single-photon wavepacket, depending on only two parameters, the superradiant enhancement $\chi$ and the Rabi frequency $\Omega_0$. Fitting Eq.~\eqref{p1} to the data on Fig.~\ref{circ01}(a), we then obtain $\chi = 4.0$ and $\Omega_0 = 0.4\times 10^9$~rad/s. These theoretical parameters would correspond to $N \approx 1.1 \times 10^6$ and $P_R = 2.1$~mW~\cite{PhysRevA.90.023848}, respectively, lying within a factor of two of our experimental estimation for these parameters. The adjustment of these quantities to the data is almost independent of each other, since the Rabi frequency is adjusted to match the observed oscillation period and $\chi$ is adjusted to match the observed decay envelop, directly shown on the blue dashed lines of Fig.~\ref{circ01}. The observation of $\chi > 1$ demonstrates then the superradiant acceleration for both single- and two-photon wavepackets.

For the cigar-shaped geometry of our atomic ensemble, we calculated in Eq.~\eqref{chi} that $\chi = 1 + N/2 w_0^2 k_{ge}^2$. Since $OD \propto N$~\cite{PhysRevA.90.023848}, we have that $\chi - 1 \propto OD$. The difference in $OD$ between Figs.~\ref{circ01}(a) and~\ref{circ01}(b) provides then directly the expected value of $\chi = 2.52$ for Fig.~\ref{circ01}(b), with no need for a new adjustment of the parameters, since $OD$ is the only experimental parameter modified between the two curves. As expected, the resulting theoretical curve on Fig.~\ref{circ01}(b) fits as well the experimental data as in Fig.~\ref{circ01}(a).

When compared to our previous investigation of the single-photon wavepacket on Ref.~\cite{PhysRevA.90.023848}, our present results show a worse comparison between experiment and theory. The main difference is the plateau on the experimental data that appear for long times. We believe this comes from higher background noise in our present case, due to the removal of a frequency filter for field $1$ employed in Ref.~\cite{PhysRevA.90.023848} and not here. In our present investigation, the removal of this filter was essential to improve the four-photon coincidence rate to acceptable levels. Another difference comes from the dip occurring in the experimental wavepackets around $t = 7$~ns. This dip has no relation to the underlying dynamics and does not move its position with any modification of the experimental parameters. It comes from an analogous dip at the beginning of the optical read pulse that we were unable to eliminate.

\begin{figure}[ht]
	\centering 
	\includegraphics[width=9.0cm]{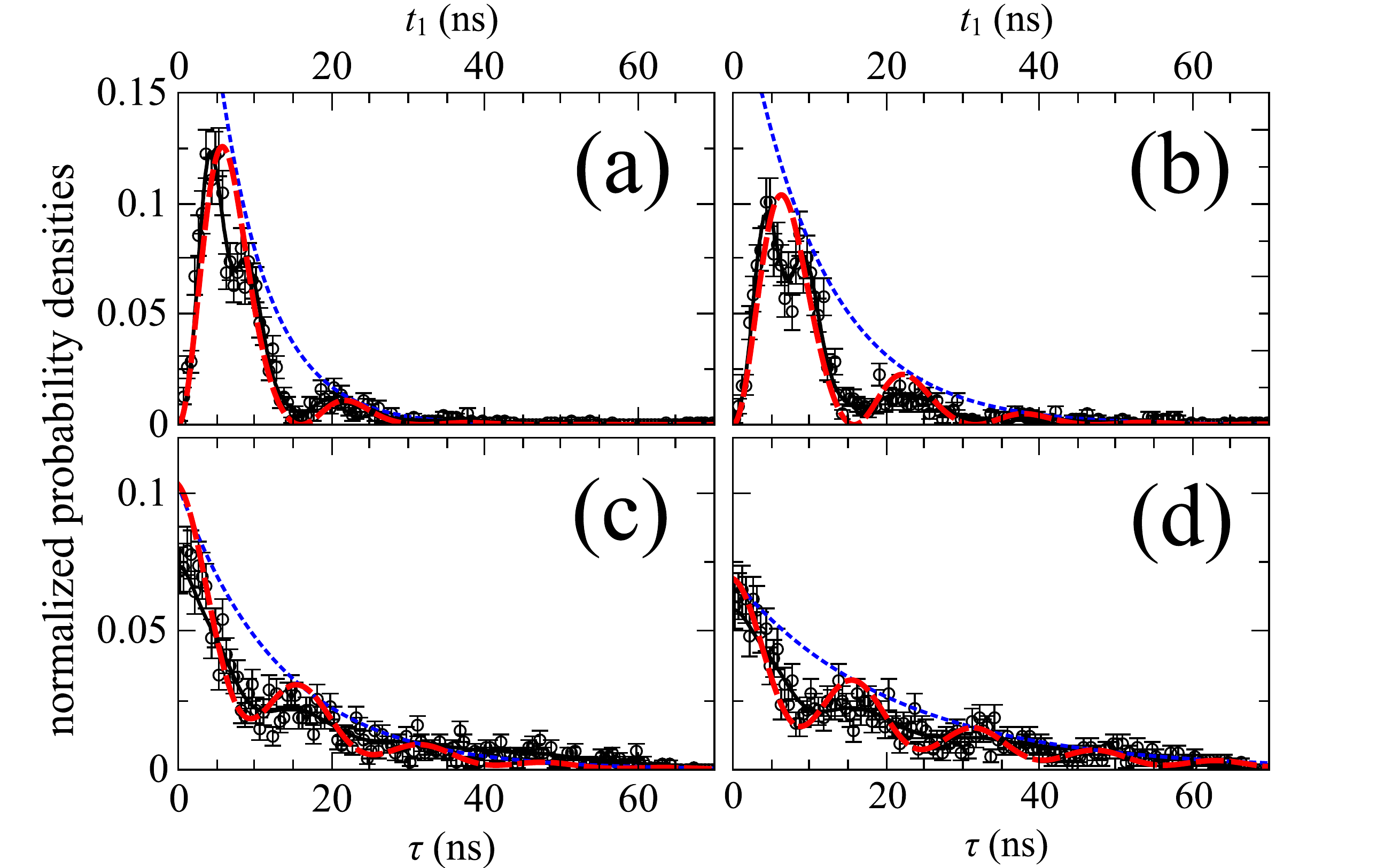}
	\caption{
		\label{circ02}
		Two-photon wavepackets for circular polarizations and optical pumping (black circles). Panels (a) and (b): Normalized probability density $\rho^{(2,2)}_{1} (t_1)$ to detect the first field-2 photon at time $t_1$. Panels (c) and (d): Normalized probability density $\rho^{(2,2)}_2 (\tau)$ to detect the second field-2 photon at time $\tau$ after the first detection. Data on panels (a) and (c) [(b) and (d)] resulted from the same measurements as for the experimental continuous black curve on Fig.~\ref{circ01}(a) [Fig.~\ref{circ01}(b)]. Traced red lines provide the theory from Eqs.~\eqref{p21} and~\eqref{second}, for the same parameters of the red curves in Fig.~\ref{circ01}. Dashed blue lines plot the respective pure exponential decays. Black continuous lines plot the respective numerical two-photon wavepackets.}
\end{figure}

Even though, the present investigation involves the comparison of experimental and theoretical wavepackets over a more significant number of curves for the same parameters, globally reinforcing the obtained fitting. For the situations of Fig.~\ref{circ01}, we plot in Fig.~\ref{circ02} the full two-photon wavepackets, divided into $\rho^{(2,2)}_1(t_1)$ and $\rho^{(2,2)}_2(\tau)$. We observe a reasonable agreement of the data to the theoretical curves coming from Eqs.~\eqref{p21} and~\eqref{second}, respectively, for the same parameters of Fig.~\ref{circ01}. As seen in the theoretical section, the curves for $\rho^{(2,2)}_1(t_1)$ are particularly suited to probe the superradiant decay, since their envelopes of the Rabi oscillations decay with $2\chi\Gamma$ instead of the $\chi \Gamma$ of the usual single-photon wavepacket. As in the case for linear polarizations, we also plot on Fig.~\ref{circ02} the figures (continuous black lines) expected for $\rho^{(2,2)}_1(t_1)$ and $\rho^{(2,2)}_2(\tau)$ from the single-photon wavepacket combined with the condition for independent emissions given by Eq.~\eqref{ind}. They follow as well the corresponding two-photon probability distribution. The small deviations for $\tau \approx 0$ have now even less significance.

\section{Conclusions}
	\label{conclusions}

With our simplified theory for the reading process of a quantum memory, we arrived at analytical expressions for the wavepacket of the extracted photon when one excitation is stored on the memory, and for the wavepackets of both extracted photons when two excitations are stored on the memory. The main difference of this theory compared to other theories for the process \cite{PhysRevLett.84.5094, RevModPhys.75.457, PhysRevA.73.021803}, that can be solved only numerically, is that we assume that we have an electromagnetically induced transparency in the medium induced by the read laser beam, and remove this complicated effect from the treatment. With some other approximations, the treatment becomes a combination of Rabi oscillations induced by the read beam between levels $\ket*{s}$ and $\ket*{e}$ on the atoms that store the excitations, no matter which atoms are these, with a superradiant spontaneous decay from level $\ket*{e}$ to level $\ket*{g}$. Superradiance appears due to the fundamental indistinguishability of which atoms store the excitations, and the coherent superposition of the different probability amplitudes lead to an increase in the decay rate. The theory thus provides a simple physical picture of the phenomena. 

Another advantage of the developed theory is that experimental curves could be fitted with only two parameters: one related to the Rabi oscillation frequency and another related to the superradiant increase of the decay rate. The same parameters were used in the experiments of Sec.~\ref{circular} to fit the single emission cases as well as the first and second photon emissions in the double emission cases, with two different optical depths (and consequently two different number of atoms) of the atomic memory. The agreement with the experimental curves demonstrates that the simplified theory kept the main physical aspects of the phenomenon.

An important prediction of the theory is that, in the two-photon emission case, both photons are emitted in the same spatiotemporal mode. This prediction may be of relevance for quantum information applications since the quantum memory can act as a superradiant source of a Fock state of light with two excitations. Even though our theory for the form of the photons wavepackets could not be used to model the experiments of Sec.~\ref{linear}, since due to the absence of optical pumping the approximation of the atoms by three-level systems was not adequate, the prediction of independent photon emissions was successfully confirmed by the experiments. The experiments of Sec.~\ref{circular} also confirmed it.

The theoretical and experimental results shown here and in the previous works from our group on the subject~\cite{1367-2630-15-7-075030, PhysRevA.90.023848, PhysRevLett.120.083603} demonstrate the power of the atomic memories based on the DLCZ protocol as convenient tools for fundamental studies of superradiance at Fock state regimes.

This work was supported by CNPq, CAPES, FAPEMIG and FACEPE (Brazilian agencies), through the programs PRONEX and INCT-IQ (Instituto Nacional de Ci\^encia e Tecnologia de Informa{\c c}\~ao Qu\^antica). C. A. E. G. acknowledges the support from the Universidad de Antioquia, Colombia (contract CODI-251594 and \textit{Estrategia de sostenibilidad del Grupo de F\'{\i}sica At\'omica y Molecular}) and Colciencias for a Ph.D. grant Cr\'edito Condonable.

\section*{Appendix: Calculus of the superradiance enhancement $\chi$}
	\label{appendix}

To obtain an expression for the superradiance enhancement $\chi$ in terms of experimentally accessible parameters, we substitute $\Phi(\vb{k})$ from Eq.~\eqref{phi} in Eq.~\eqref{fatorsuper}:
\begin{equation}
	\label{chi-a}
	\chi=\frac{1}{4\pi k_{ge}^2} \sum_{ij} \alpha_i \alpha_j^* \oint_{\abs{\vb{k}} = k_{ge}} \dd[2]{\vb{k}} \me^{\mi(\vb{k}_\text{r} - \vb{k}) \cdot (\vb{r}_i - \vb{r}_j)}.
\end{equation}
For every $i = j$ term in this double sum, we have the normalization condition $\sum_i \abs{\alpha_i}^2 = 1$ and the integral results in $4\pi k_{ge}^2$, such that these terms with $i = j$ contribute with a term $1$ for $\chi$. To compute the remaining terms of the summations, we use the fact that the photon emitted in the read process of the memory is highly directional, being emitted in the opposite direction of the photon detected in the write process, as evidenced by Eq. \eqref{phi}. This allows the simplification of the integration region from the sphere $\abs{\vb{k}} = k_{ge}$ into a small cap $D$ around $-\vb{k}_1$, where $\vb{k}_1$ is the central wavevector of the photon detected in the write process of the memory. For the other directions, the summation over the atoms positions in Eq. \eqref{phi} results in destructive interference, with a negligible contribution to $\chi$. Around this small region $D$, the contribution of the $i = j$ terms is negligible compared to 1, since the small cap $D$ is much smaller than the sphere surface. With these approximations, we can write Eq.~\eqref{fatorsuper} as
\begin{equation}
	\label{phi2}
	\chi\approx 1 + \frac{1}{4\pi k_{ge}^2} \int_D \dd[2]{\vb{k}} \abs{\Phi(\vb{k})}^2,
\end{equation}
where the first term comes from the $i = j$ terms in Eq.~\eqref{chi-a} and the second term from the $i \neq j$ terms, considering that the contribution of the $i = j$ terms present in this integral is much smaller than one. 

As it was discussed after Eq.~\eqref{phi}, we have $\Phi(\vb{k}) \propto \phi_1(-\vb{k})$, where $\phi_1(\vb{k})$ is the mode of the photon detected in the write process. In our experiments, this is a paraxial Gaussian mode
\begin{equation}
	\phi_1(k_x,k_y) \propto \me^{-(k_x^2+k_y^2)w_0^2/4},
\end{equation}
with the $z$ axis is defined in the direction of $\vb{k}_1$. The amplitude of the field in the center of the atomic cloud is then a Gaussian $E_0 \me^{-(x^2+y^2)/w_0^2}$. So we can write
\begin{equation}
	\label{phiap}
	\Phi(k_x,k_y) = C \me^{-(k_x^2+k_y^2)w_0^2/4},
\end{equation}
where $C$ is a constant and the $z$ axis is defined in the direction of $-\vb{k}_1$ in this case. Substituting this expression for $\Phi(\vb{k})$ in Eq.~\eqref{phi2}, we obtain
\begin{equation}
\label{phi3}
\chi\approx 1+ \frac{\abs{C}^2}{2w_0^2 k_{ge}^2}.
\end{equation}
The constant $\abs{C}$ can be obtained by considering that the maximum value for $\abs{\Phi(\vb{k})}$ is obtained when the scattering by all atoms interfere constructively to the photon emission in a particular direction. According to Eqs.~\eqref{phiap} and \eqref{phi}, this can be written as $\abs{\Phi(-\vb{k}_1)} = \abs{C} = \sum_i \abs{\alpha_i}$. To give an approximate value for this constant, let us make the following approximation. Instead of considering all atoms $N'$ that interact with the mode detected in the write process (mode 1), let us consider an effective number of atoms $N$. On the approximation, we substitute the $N'$ atoms interacting with mode 1 with coupling constants proportional to $\alpha_i$, which depends on the atom position due to the Gaussian form of the mode, by $N$ atoms that interact with the mode with the same coupling proportional to $\alpha_0$, which is the amplitude associated to the atoms in the center of the mode. With this consideration, the normalization condition $\sum_i \abs{\alpha_i}^2 = N \abs{\alpha_0}^2 = 1$ implies that $\alpha_0 = 1/\sqrt{N}$. So we have $\abs{C} = N \alpha_0 = \sqrt{N}$. Substituting this value in Eq.~\eqref{phi3}, we obtain Eq.~\eqref{chi}.

\end{document}